\documentclass[pre,onecolumn,aps,nofootinbib]{revtex4}

\usepackage{amsfonts}
\usepackage{amsmath}
\usepackage{graphicx}        % standard LaTeX graphics tool
\newcommand{\rmpp}{{\mathrm{p}}}
\newcommand{\rmd}{{\mathrm{d}}}
\newcommand{\rmi}{{\mathrm{i}}}

\begin{document}

\title{Basic microscopic plasma physics unified and simplified by $N$-body classical mechanics
V4 - 22/11/2012}

\author{D. F. Escande, F. Doveil, and Y. Elskens}

\affiliation{UMR 7345 CNRS--Aix-Marseille-Universit\'e, Facult\'es
de Saint J\'er\^ome, case 321, \\ Av. Normandie Niemen, FR-13397 Marseille CEDEX 20}
%\email{Dominique.Escande@univ-amu.fr}

\begin{abstract}
In typical plasma physics textbooks, Debye shielding, collisional transport, Landau damping of Langmuir waves, and spontaneous emission of these waves are introduced in different chapters. This paper provides a compact unified introduction to these phenomena without appealing to fluid or kinetic models, but by using Newton's second law for a system of $N$ electrons in a periodic box with a neutralizing ionic background. A rigorous equation is derived for the electrostatic potential. Its linearization and a first smoothing reveal this potential to be the sum of the shielded Coulomb potentials of the individual particles. Smoothing this sum yields the classical Vlasovian expression including initial conditions in Landau contour calculations of Langmuir wave growth or damping. The theory is extended to accommodate a correct description of trapping or chaos due to Langmuir waves. In the linear regime, the amplitude of such a wave is found to be ruled by Landau growth or damping and by spontaneous emission. Using the shielded potential, the collisional diffusion coefficient is computed for the first time by a convergent expression including the correct calculation of deflections for all impact parameters. Shielding and collisional transport are found to be two related aspects of the repulsive deflections of electrons.

\end{abstract}

\maketitle

%VVVVVVVVVVVVVVVVVVVVVVVVVVVVVVVVVVVVVVV
\section{Introduction}
\label{sec:1}
%VVVVVVVVVVVVVVVVVVVVVVVVVVVVVVVVVVVVVVV

From the outset, inspired by gas physics, plasma physicists derived kinetic equations to describe microscopic aspects of their physics, in particular the Vlasov equation. This trend has been the dominant one till nowadays. However, for plasmas where transport due to short range interactions (``collisions") is weak, it is possible to work directly with $N$-body classical mechanics. As will be recalled in section \ref{WPDRT}, this approach led to a new description of wave-particle interaction making it more intuitive and unifying particle and wave evolutions, as well as collective and finite-$N$  physics.

This paper develops this approach further by deriving a rigorous fundamental equation for the electrostatic potential (section \ref{FEP}), which is extended to enable a correct description of trapping or chaos due to Langmuir waves (section \ref{WPDRT}). It brings further unifications: Debye shielding (or screening) and Landau theory of linear waves (section \ref{DSLD}), Landau growth or damping and spontaneous emission (section \ref{WPDRT}), Debye shielding and collisional transport (section \ref{DSCT}). These results come with a considerable simplification of the mathematical framework with respect to textbooks (see for instance Refs. \cite{HW,Nic,Chen,BS}), and with new insights into microscopic plasma physics. In particular it solves the following issue: How can each particle be shielded by all other ones, while all the plasma particles are in uninterrupted motion? This turns out to be a mere consequence of the almost independent deflections of particles due to the Coulomb force. Furthermore, the new path to Landau damping goes first through Debye shielding, a totally unexpected fact, since classical textbooks present these concepts in different and unrelated chapters.

As to collisional transport, this paper does not simply recover the classical result presented in textbooks, but it provides the first calculation of this transport without any ad hoc cut-off. This is made possible by a mechanical calculation of the deflection of a particle which covers the scales corresponding to the Debye length, to the inter-particle distance, and to the distance of minimum approach of two electrons in a Rutherford collision. Till now none of the papers about collisional transport, and no textbook, was able to correctly describe the scale of the inter-particle distance.

The derivations are elementary (we refrain from invoking the Liouville equation or the BBGKY hierarchy). After section \ref{FEP}, the successive approximations of the fundamental equation (\ref{phihat}) enable to get rid of the heavy notations related to its generality. These approximations involve linearization in sections \ref{FEP}, \ref{SCP}, and \ref{STSP}, but section \ref{MIIDS}, the beginning of section \ref{WPDRT}, and section \ref{DSCT} do not appeal to linearization. For the sake of brevity, this paper is more oriented toward concepts and intuitive physics, than toward ready-to-use formulas to be presented elsewhere.

%VVVVVVVVVVVVVVVVVVVVVVVVVVVVVVVVVVVVVVV
\section{Claims}
\label{Cl}
%VVVVVVVVVVVVVVVVVVVVVVVVVVVVVVVVVVVVVVV

Here are the main results of this paper:

1. By using the Fourier and Laplace transforms, a rigorous equation (Eq.~(\ref{phihat})) is derived for the electrostatic potential of an infinite plasma made up of the periodic replication of $N$ electrons coupled by Coulomb forces in a volume $L^3$ with a neutralizing ionic background (One Component Plasma (OCP) model \cite{Salp,Abe,BH}). This equation is of the type ${\mathcal{E}}\hat \varphi=$ source term, where ${\mathcal{E}}$ is a linear operator, acting on the infinite dimensional array $\hat \varphi$ whose components are all the Fourier-Laplace components of the potential.

2. Considering that particle trajectories stay close to ballistic, and replacing inside ${\mathcal{E}}$ the discrete sums over particles by integrals over a smooth distribution function, the potential turns out to be the sum of the shielded Coulomb potentials of the individual particles (Eq.~(\ref{phi})). Such potentials were first computed by a kinetic approach in section II.A of Ref.\ \cite{Gasio} and later on in \cite{Bal,Rost}.

3. Replacing the discrete sums over particles of their shielded potentials by integrals over a smooth distribution function, yields the classical Vlasovian expression (Eqs~(\ref{phihatL}) and (\ref{phi0hatcg})) including initial conditions in Landau contour calculations of Langmuir wave growth or damping, obtained by linearizing Vlasov equation and using Fourier-Laplace transform, as described in many textbooks (see for instance Refs. \cite{HW,Nic,BS}). Therefore in these calculations, the electrostatic potential turns out to be the smoothed version of the actual shielded potential in the plasma.

4. In the spirit of Refs. \cite{OWM,OLMSS,AEE,EEB}, in order to accommodate a correct description of trapping or chaos due to Langmuir waves, the set of particles is split into bulk and tail, where the bulk is the set of particles which cannot resonate with Langmuir waves. Repeating the analysis leading to Eq.~(\ref{phihat}) for the bulk particles, the same equation is recovered with an additional source term due to the tail particles (Eq.~(\ref{phihatU})).

5. Using the fact that the number of tail particles is small with respect to the bulk one, and a technique introduced in Refs. \cite{OWM,OLMSS}, an amplitude equation is derived for any Fourier component of the potential where tail particles provide a source term (Eq.~(\ref{eqampl})).

6. This equation, together with the equation of motion of the tail particles, enables to show that, in the linear regime, the amplitude of a Langmuir wave is ruled by Landau growth or damping, and by spontaneous emission (Eq.~(\ref{evampfinal})), a generalization to 3 dimensions of a one-dimensional result of Refs. \cite{EZE,EEB}.

7. Using the shielded potential, the collisional diffusion coefficient $D$ of a given particle is computed by a convergent expression including the particle deflections for all impact parameters. These deflections are computed by first order perturbation theory in the total electric field, except for those due to close encounters. The contribution to $D$ of the former ones is matched with that of the latter ones provided by Ref.\ \cite{Ros}. The detailed matching procedure includes the scale of the inter-particle distance, and is reminiscent of that in \cite{Hub}, without invoking the cancellation of three infinite integrals. $D$ has the same expression as that in Ref.\ \cite{Ros}, except for the Coulomb logarithm which is modified by a velocity dependent quantity of order 1.

8. Picard iteration technique is applied to the equation of motion of particle $P$ due to the Coulomb forces of all other ones. It reveals that a part of the effect on particle $P$ of another particle $P'$ is mediated by all other particles (Eq.~(\ref{rsecnAccDev2})). Indeed particle $P'$ modifies the position of all other particles, implying that the action of the latter ones on particle $P$ is modified by particle $P'$.

9. This calculation yields the following interpretation of shielding.
At $t =0$ consider a set of randomly distributed particles, and especially particle $P$. At a later time $t$, the latter has deflected all particles which made a closest approach to it with an impact parameter $b \lesssim v_{\rm{th}} t$ where $v_{\rm{\rm{th}}}$ is the thermal velocity. This part of their global deflection due to particle $P$ reduces the number of particles inside the sphere $S(t)$ of radius $v_{\rm{th}} t$ about it. Therefore the effective charge of particle $P$ as seen out of $S(t)$ is reduced: the charge of particle $P$ is shielded due to these deflections. This shielding effect increases with $t$, and thus with the distance to particle $P$. As a result, the typical time-scale for shielding to set in when starting from random particle positions is the time for a thermal particle to cross a Debye sphere, i.e.\ $\omega_{\rmpp}^{-1}$,  where $\omega_{\rmpp}$ is the plasma frequency. Furthermore, shielding is a cooperative dynamical process: it results from the accumulation of almost independent repulsive deflections with the same qualitative impact on the effective electric field of particle $P$ (if point-like ions were present, the attractive deflection of charges with opposite signs would have the same effect). It is a cooperative effect, but not a collective one. So, shielding and collisional transport are two aspects of the same two-body repulsive process.

%VVVVVVVVVVVVVVVVVVVVVVVVVVVVVVVVVVVVVVV
\section{Fundamental equation for the potential}
\label{FEP}
%VVVVVVVVVVVVVVVVVVVVVVVVVVVVVVVVVVVVVVV

This paper deals with the One Component Plasma model \cite{Salp,Abe,BH}, which considers the plasma as infinite with spatial periodicity $L$ in three orthogonal directions with coordinates $(x,y,z)$, and as made up of $N$ electrons in each elementary cube with volume $L^3$. Ions are present only as a uniform neutralizing background, which enables periodic boundary conditions. This choice is made to simplify the analysis which focuses on $\varphi(\textbf{r})$, the potential created by the $N$ particles at any point where there is no particle. The discrete Fourier transform of $\varphi(\textbf{r})$, readily obtained from the Poisson equation, is given by $\tilde{\varphi}(\textbf{0}) = 0$ and, for $\textbf{m} \neq \textbf{0}$, by
\begin{equation}
  \tilde{\varphi}(\textbf{m})
  = - \frac{e}{\varepsilon_0 k_{\textbf{m}}^2} \sum_{j \in S}
      \exp(- \rmi \textbf{k}_{\textbf{m}} \cdot \textbf{r}_j),
\label{phitildetotM}
\end{equation}
where $-e$ is the electron charge, $\varepsilon_0$ is the vacuum permittivity, $\textbf{r}_j$ is the position of particle $j$, $S$ is the set of integers going from 1 to $N$, $\tilde{\varphi}(\textbf{m}) = \int \varphi(\textbf{r}) \exp(- \rmi \textbf{k}_{\textbf{m}} \cdot \textbf{r}) \rmd^3 \textbf{r}$, $\textbf{m} = (m_x,m_y,m_z)$ is a vector with three integer components,  $\textbf{k}_{\textbf{m}} = \frac{2 \pi}{L}\textbf{m}$, and $k_{\textbf{m}} = \|\textbf{k}_{\textbf{m}}\|$. Reciprocally,
\begin{equation}
\varphi(\textbf{r}) = \frac{1}{L^3}\sum_{\textbf{m}} \tilde{\varphi} (\textbf{m}) \exp(\rmi \textbf{k}_{\textbf{m}} \cdot \textbf{r}),
\label{phiInv}
\end{equation}
where $\sum_{\textbf{m}}$ means the sum over all components of $\textbf{m}$ running from $- \infty$ to $+ \infty$.

The dynamics of particle $l$ is determined by Newton's equation
\begin{equation}
\ddot{\textbf{r}}_l = \frac{e}{m_e} \nabla \varphi_l(\textbf{r}_l),
\label{rsectot}
\end{equation}
where $m_e$ is the electron mass, $\varphi_l$ is the electrostatic potential acting on particle $l$, i.e.\ the one created by all other particles and by
the background charge. Its Fourier transform is given by Eq.~(\ref{phitildetotM}) with the supplementary condition $j \neq l$.
Let
\begin{equation}
\textbf{r}_l^{(0)} = \textbf{r}_{l0} + \textbf{v}_{l} t
\label{rl0}
\end{equation}
be a ballistic approximation of the motion of particle $l$. In the following, we consider two instances of this approximation: the one where $\textbf{r}_{l0}$ and $\textbf{v}_{l}$ are respectively the initial position and velocity of particle $l$, and the one where they are slightly shifted from these values by low amplitude Langmuir waves. Let $\delta \textbf{r}_l = \textbf{r}_l - \textbf{r}_l^{(0)}$, so that Eq.~(\ref{rsectot}) is equivalent to
\begin{equation}
\delta \ddot{\textbf{r}}_l = \frac{\rmi e}{L^3 m_e} \sum_{\textbf{n}} \textbf{k}_{\textbf{n}} \ \tilde{\varphi}_l(\textbf{n}) \exp[\rmi \textbf{k}_{\textbf{n}} \cdot (\textbf{r}_l^{(0)} + \delta \textbf{r}_l)].
\label{delrsec}
\end{equation}
Sections \ref{DSLD} and \ref{WPDRT} consider cases where the $\delta \textbf{r}_l$'s are small in some sense. To this end, we split $\tilde{\varphi}_l(\textbf{m})$ as
\begin{equation}
\tilde{\varphi}_l(\textbf{m}) = \tilde{\phi}_l(\textbf{m}) + \Delta \tilde{\varphi}_l(\textbf{m})
\label{phitildnapp}
\end{equation}
where
\begin{equation}
\tilde{\phi}_l (\textbf{m}) = \sum_{j \in S;j \neq l} \delta \tilde{\phi}_{j} (\textbf{m}),
\label{phitildn}
\end{equation}
with
\begin{equation}
\delta \tilde{\phi}_{j} (\textbf{m}) = -\frac{e}{\varepsilon_0 k_{\textbf{m}}^2} \exp(- \rmi \textbf{k}_{\textbf{m}} \cdot \textbf{r}_{j}^{(0)})(1 - \rmi \textbf{k}_{\textbf{m}} \cdot \delta \textbf{r}_j),
\label{phitildnj}
\end{equation}
and
\begin{equation}
\Delta \tilde{\varphi}_l(\textbf{m}) = -\frac{e}{\varepsilon_0 k_{\textbf{m}}^2} \sum_{j \in S;j \neq l}
\exp(- \rmi \textbf{k}_{\textbf{m}} \cdot \textbf{r}_{j}^{(0)}) R_j(\textbf{m}),
\label{Deltaphi}
\end{equation}
with
\begin{equation}
R_j(\textbf{m}) = \exp(- \rmi \textbf{k}_{\textbf{m}} \cdot \delta \textbf{r}_j) -1 + \rmi \textbf{k}_{\textbf{m}} \cdot \delta \textbf{r}_j,
\label{Rj}
\end{equation}
which is of order two in $\delta \textbf{r}_j$.

We now introduce the time Laplace-transform which transforms a function $ f(t)$ into $\hat{f}(\omega) = \int_0^{\infty}  f(t) \exp(\rmi \omega t) \rmd t$ (with $\omega$ complex). The Laplace transform of Eq.~(\ref{delrsec}) is
\begin{equation}
\omega^2 \delta \hat{\textbf{r}}_l(\omega) = - \frac{\rmi e}{L^3 m_e} \sum_{\textbf{n}} \textbf{k}_{\textbf{n}} \exp(\rmi \textbf{k}_{\textbf{n}} \cdot \textbf{r}_{l0}) \ \Psi_l(\hat{\varphi}_l \ ; \textbf{n},\omega + \omega_{\textbf{n},l}) + \rmi \omega \delta \textbf{r}_l(0) - \delta \dot{\textbf{r}}_l(0).
\label{rLapl}
\end{equation}
where carets indicate the Laplace transformed versions of the quantities in Eq.~(\ref{delrsec}), $\omega_{\textbf{n},l} = \textbf{k}_{\textbf{n}} \cdot \textbf{v}_{l}$ comes from the time dependence of $\textbf{r}_l^{(0)}$ in the exponent of Eq.~(\ref{delrsec}), and the operator $\Psi_l$ acting on a function $g(\textbf{m},\omega)$ is defined by
\begin{equation}
\Psi_l(g\ ; \textbf{n},\cdot) = g(\textbf{n},\cdot) \ast T_{l}(\textbf{n},\cdot),
\label{psil}
\end{equation}
where $\cdot$ stands for the frequencies, $ \ast $ is the convolution product, and $T_{l}(\textbf{n},\omega)$ is the Laplace transform of $\exp(\rmi \textbf{k}_{\textbf{n}} \cdot \delta \textbf{r}_l)$. The Laplace transform of Eqs~(\ref{phitildnapp})-(\ref{Rj}) yields
\begin{equation}
 k_{\textbf{m}}^2\hat{\varphi}_l(\textbf{m},\omega) = k_{\textbf{m}}^2 \hat{\phi}_l^{(00)}(\textbf{m},\omega) + \frac{\rmi e}{\varepsilon_0} \sum_{j \in S;j \neq l} \exp(- \rmi \textbf{k}_{\textbf{m}} \cdot \textbf{r}_{j0}) \ [\textbf{k}_{\textbf{m}} \cdot \delta \hat{\textbf{r}}_j(\omega - \omega_{\textbf{m},j}) + \rmi \hat{R}_j(\textbf{m},\omega - \omega_{\textbf{m},j})],
\label{phihatn}
\end{equation}
where $\hat{R}_j(\textbf{m},\omega)$ is the Laplace transform of $R_j$, and
% $\omega_{\textbf{m},j} = \textbf{k}_{\textbf{m}} \cdot \textbf{v}_{j}$
% comes from the time dependence of $\textbf{r}_l^{(0)}$
% in the exponent of Eqs~(\ref{phitildnj})-(\ref{Deltaphi});
$\hat{\phi}_l^{(00)}(\textbf{m},\omega) $ is the Laplace transform of $\tilde{\phi}_l (\textbf{m}) $ computed from Eqs~(\ref{phitildn}) and (\ref{phitildnj}) by setting $\delta \textbf{r}_j =0$ for all $j$'s in the latter. Substituting the $\delta \hat{\textbf{r}}_j$'s with their expression Eq.~(\ref{rLapl}) yields
\begin{eqnarray}
  &&
  k_{\textbf{m}}^2 \hat{\varphi}_l(\textbf{m},\omega)
  - \frac{e^2}{ L^3 m_e \varepsilon_0}
  \sum_{\textbf{n}} \textbf{k}_{\textbf{m}} \cdot \textbf{k}_{\textbf{n}}
  \ \sum_{j \in S;j \neq l}
     \frac{\Psi_j(\hat{\varphi}_j\ ; \textbf{n},\omega + \omega_{\textbf{n},j} - \omega_{\textbf{m},j})}
          {(\omega - \omega_{\textbf{m},j})^2}
      \exp[\rmi (\textbf{k}_{\textbf{n}}-\textbf{k}_{\textbf{m}})
           \cdot \textbf{r}_{j0}]
  \nonumber\\
  & = &
  k_{\textbf{m}}^2 \hat{\phi}_l^{(0)}(\textbf{m},\omega)
  - \frac{e}{\varepsilon_0} \sum_{j \in S;j \neq l}
    \exp(- \rmi \textbf{k}_{\textbf{m}} \cdot \textbf{r}_{j0})
    \hat{R}_j(\omega - \omega_{\textbf{m},j}),
\label{phihatnf}
\end{eqnarray}
where $\hat{\phi}_l^{(0)}(\textbf{m},\omega) $ is the Laplace transform of $\tilde{\phi}_l (\textbf{m}) $ computed from Eqs~(\ref{phitildn}) and (\ref{phitildnj}) by setting now $\delta \textbf{r}_j = \delta \textbf{r}_j(0) + \delta \dot{\textbf{r}}_j(0) t$ for all $j$'s in the latter.

Summing Eq.~(\ref{phihatnf}) over $l = 1,... N$ and dividing by $N-1$, yields
\begin{eqnarray}
 k_{\textbf{m}}^2\hat{\varphi}(\textbf{m},\omega)
&-& \frac{e^2}{ L^3 m_e \varepsilon_0}
 \sum_{\textbf{n}} \textbf{k}_{\textbf{m}} \cdot \textbf{k}_{\textbf{n}}
\ \sum_{j \in S} \frac{\Psi_j(\hat{\varphi}\ ; \textbf{n},\omega + \omega_{\textbf{n},j} - \omega_{\textbf{m},j})}{(\omega - \omega_{\textbf{m},j})^2} \exp[\rmi (\textbf{k}_{\textbf{n}}-\textbf{k}_{\textbf{m}}) \cdot \textbf{r}_{j0}]
\nonumber\\
= k_{\textbf{m}}^2 \hat{\phi}^{(0)}(\textbf{m},\omega) &-& \frac{e}{\varepsilon_0} \sum_{j \in S} \exp(- \rmi \textbf{k}_{\textbf{m}} \cdot \textbf{r}_{j0}) \hat{R}_j(\omega - \omega_{\textbf{m},j}),
\label{phihat}
\end{eqnarray}
where $\hat{\phi}^{(0)}(\textbf{m},\omega) $ is $\hat{\phi}_l^{(0)}(\textbf{m},\omega) $ complemented by the missing $l$-th term.
Equation (\ref{phihat}) is the fundamental equation of this paper, and is a rigorous consequence of Eqs~(\ref{phitildetotM}) and (\ref{rsectot}): no approximation was made. This fundamental equation is of the type ${\mathcal{E}}\hat \varphi=$ source term, where ${\mathcal{E}}$ is a linear operator, acting on the infinite dimensional array whose components are all the $\hat{\varphi}(\textbf{m},\omega)$'s.

%VVVVVVVVVVVVVVVVVVVVVVVVVVVVVVVVVVVVVVV
\section{Debye shielding}
\label{DSLD}
%VVVVVVVVVVVVVVVVVVVVVVVVVVVVVVVVVVVVVVV

%VVVVVVVVVVVVVVVVVVVVVVVVVVVVVVVVVVVVVVV
\subsection{Shielded Coulomb potential}
\label{SCP}
%VVVVVVVVVVVVVVVVVVVVVVVVVVVVVVVVVVVVVVV

We now specialize Eq.~(\ref{phihat}) by considering the lowest order contribution of the $\delta \textbf{r}_j$'s, which makes the $\hat{R}_j$'s vanish  (Approximation 1). We further consider $\varphi$ to be small, and the $\delta \textbf{r}_l$'s to be of the order of $\varphi$ (Approximation 2). These two approximations reduce $\Psi_j(\hat \varphi ; \textbf{n},\omega)$ to $\hat{\varphi}(\textbf{n},\omega)$. This corresponds to substituting the true dynamics in Eq.~(\ref{rsectot}) with an approximate one ruled by
\begin{equation}
\delta \ddot{\textbf{r}}_l = \frac{e}{m_e} \nabla \phi_l(\textbf{r}_l^{(0)} + \delta \textbf{r}_l),
\label{rsec}
\end{equation}
where $\phi_l (\textbf{r}) = \sum_{j \in S;j \neq l} \delta \phi_{j} (\textbf{r})$ is the inverse Fourier transform of Eq.~(\ref{phitildn}), so that
\begin{equation}
\lim_{L \to \infty} \delta \phi_{j} (\textbf{r}) = - \frac{e}{4 \pi \varepsilon_0 \| \textbf{r} - \textbf{r}_j^{(0)}\|} - \frac{e \delta \textbf{r}_j \cdot (\textbf{r} - \textbf{r}_j^{(0)})}{4 \pi \varepsilon_0 \| \textbf{r} - \textbf{r}_j^{(0)}\|^3},
\label{deltaphi}
\end{equation}
The $j$th component of the approximate electric field acting over particle $l$ turns out to be due to a particle located at $\textbf{r}_{j}^{(0)}$ instead of $\textbf{r}_j$, and is made up of a Coulombian part and of a dipolar part with a dipole moment $- e \delta \textbf{r}_j$. The cross-over between the two contributions occurs for $\| \textbf{r}_l - \textbf{r}_j^{(0)}\|$ on the order of $\| \delta \textbf{r}_j \|$, i.e.\ when the distance between particle $l$ and the ballistic particle $j$ is equal to the distance between the latter and the true particle $j$. For larger values of $\| \textbf{r}_l - \textbf{r}_j^{(0)}\|$, the dipolar component is subdominant. For smaller ones, it is dominant, but with a direction which is a priori random with respect to the Coulombian one ($(\textbf{r}_l - \textbf{r}_j^{(0)})$ is almost independent from $\delta \textbf{r}_j$). Since the $\| \delta \textbf{r}_j \|$'s are assumed small, the latter case should be rare since it corresponds to a very close encounter between particle $l$ and the ballistic particle $j$. As a result the approximate electric field stays dominantly of Coulombian nature, but with a small mismatch of the charge positions with respect to the actual ones.

We introduce a smooth function $f(\textbf{r},\textbf{v})$,
the smoothed velocity distribution function at $t=0$ such that
the distribution
\begin{equation}
    \sum_{l \in S} \bullet
    =
    \iint  \bullet  f(\textbf{r},\textbf{v}) \rmd^3 \textbf{r} \, \rmd^3 \textbf{v} + W(\bullet),
\label{fxv}
\end{equation}
where the distribution $W$ yields a negligible contribution when applied to space dependent function which evolve slowly on the scale of the inter-particle distance; there the spatial integration is performed over the elementary cube with volume $L^3$, and the velocity integration is over all velocities.

Replacing the discrete sums over particles by integrals over the smooth distribution function $f(\textbf{r},\textbf{v})$ (Approximation 3), Eq.~(\ref{phihat}) becomes
\begin{equation}
k_{\textbf{m}}^2 \hat{\Phi}(\textbf{m},\omega)
= k_{\textbf{m}}^2 \hat{\phi}^{(0)}(\textbf{m},\omega)
+ \frac{e^2}{L^3 m_e \varepsilon_0} \sum_{\textbf{n}} \textbf{k}_{\textbf{m}} \cdot \textbf{k}_{\textbf{n}}
\int  \frac{\hat{\Phi}(\textbf{n},\omega + (\textbf{k}_{\textbf{n}} - \textbf{k}_{\textbf{m}}) \cdot \textbf{v})}{(\omega -\textbf{k}_{\textbf{m}} \cdot \textbf{v})^2} \tilde{f}({\textbf{n}} - {\textbf{m}},\textbf{v}) \ \rmd^3 \textbf{v},
\label{phihatcg}
\end{equation}
where $\hat{\Phi}$ is the smoothed version of $\hat{\varphi}$ resulting from Approximations 1 to 3, and $\tilde{f}$ is the spatial Fourier transform of $f$. We further assume the initial distribution $f$ to be a spatially uniform distribution function $f_0(\textbf{v})$ plus a small perturbation of the order of $\Phi$ (in agreement with Approximation 2). Then operator $\mathcal{E}$ becomes diagonal with respect to both $\textbf{m}$ and $\omega$ (a complex quantity), and linearizing Eq.~(\ref{phihatcg}) for $\hat{\Phi}$ yields
\begin{equation}
\epsilon(\textbf{m},\omega) \hat{\Phi}(\textbf{m},\omega) = \hat{\phi}^{(0)}(\textbf{m},\omega),
\label{phihatL}
\end{equation}
where
\begin{equation}
\epsilon(\textbf{m},\omega) = 1 - \frac{e^2}{L^3 m_e \varepsilon_0}
\int \frac{f_0(\textbf{v}) }{(\omega -\textbf{k}_{\textbf{m}}  \cdot \textbf{v})^2} \ \rmd^3 \textbf{v}.
\label{eps}
\end{equation}
This shows that the smoothed self-consistent potential $\hat{\Phi}$ is determined by the response function $\epsilon(\textbf{m},\omega)$. It is the classical plasma dielectric function. A first check of this can be obtained for a cold plasma: then
$\epsilon(\textbf{m},\omega) = 1 - {\omega_{\rmpp}^2}/{\omega^2}$,
where $\omega_{\rmpp}^2 = [(e^2 n)/(m_e \epsilon_0)]^{1/2}$ is the plasma frequency ($n = N/L^3$ is the plasma density). The classical expression involving the gradient of $f_0$ in $\textbf{v}$ is obtained by a mere integration by parts.

The contribution of particle $j$ to $\tilde{\phi}^{(0)}(\textbf{m})$ is $\delta \tilde{\phi}_j^{(0)}(\textbf{m}) = - \frac{e}{\varepsilon_0 k_{\textbf{m}}^2} \exp[- \rmi \textbf{k}_{\textbf{m}}  \cdot (\textbf{r}_{j0} + \textbf{v}_j t)]$. Its Laplace transform is
\begin{equation}
\delta \hat{\phi}_j^{(0)}(\textbf{m},\omega) = - \frac{\rmi e}{\varepsilon_0 k_{\textbf{m}}^2}
\frac{\exp(- \rmi \textbf{k}_{\textbf{m}}  \cdot \textbf{r}_{j0})}{\omega -\textbf{k}_{\textbf{m}}  \cdot \textbf{v}_j}.
\label{phij0hat}
\end{equation}
The corresponding part of $\hat{\Phi}(\textbf{m},\omega)$ is $\delta \hat{\Phi}_j(\textbf{m},\omega) = \delta \hat{\phi}_j^{(0)}(\textbf{m},\omega)/\epsilon(\textbf{m},\omega)$. This turns out to be the shielded potential of particle $j$ \cite{Gasio,Bal,Rost}.
By inverse Fourier-Laplace transform, after some transient whose duration is estimated at the end of section \ref{MIIDS}, the potential due to particle $j$ becomes the shielded Coulomb potential
\begin{equation}
\delta \Phi_j (\textbf{r}) = \delta \Phi(\textbf{r} - \textbf{r}_{j0} - \textbf{v}_j t,\textbf{v}_j),
\label{phij}
\end{equation}
where
\begin{equation}
\delta \Phi (\textbf{r},\textbf{v}) = - \frac{e}{L^3 \varepsilon_0} \sum_{\textbf{m}} \frac{\exp(\rmi \textbf{k}_{\textbf{m}} \cdot \textbf{r})}{ k_{\textbf{m}}^2 \epsilon(\textbf{m},\textbf{k}_{\textbf{m}} \cdot \textbf{v})}.
\label{phi}
\end{equation}
Therefore, after this transient, the dominant contribution to the full potential in the plasma turns out to be the sum of the shielded Coulomb potentials of individual particles located at their ballistic positions. Let $\lambda_{\rm{D}} = [(\epsilon_0 k_{\rm{B}} T)/(n e^2)]^{1/2}
     = [(k_{\rm{B}} T)/m_e]^{1/2} \omega_{\rmpp}^{-1}$ be the Debye length, where $k_{\rm{B}}$ is the Boltzmann constant, $T$ is the temperature, and $n$ is the density. The wavenumbers resolving scale $\| \textbf{r}\|$ are such that $k_{\textbf{m}} \|\textbf{r}\| \gtrsim 1$. If $\|\textbf{r}\| \ll \lambda_{\rm{D}}$, the corresponding wavenumbers are such that $k_{\textbf{m}} \lambda_{\rm{D}} \gg 1$. Therefore there is no shielding for $\|\textbf{r}\| \ll \lambda_{\rm{D}}$, since $\epsilon(\textbf{m},\textbf{k}_{\textbf{m}} \cdot \textbf{v}) - 1 \simeq (k_{\textbf{m}} \lambda_{\rm{D}})^{-2}$.

In the following, Eq.~(\ref{phij}) is used by substituting $\delta \Phi(\textbf{r} - \textbf{r}_{j0} - \textbf{v}_j t,\textbf{v}_j)$ with $\delta \Phi(\textbf{r} - \textbf{r}_{j},\textbf{v}_j)$: the shielded potential of particle $j$ is computed by taking into account its actual position, since it is the original Coulomb one close to $\textbf{r}_{j}$. The error made for $\textbf{r} - \textbf{r}_{j} $ of the order of $\lambda_{\rm{D}}$ is small as long as the mismatch of $\textbf{r}_{j} $ from the ballistic orbit is much smaller than $\lambda_{\rm{D}}$. As was done for the bare potential of Eq.~(\ref{phitildetotM}), the field acting on a given particle $l$ is obtained by removing its own divergent contribution $\delta \Phi_l$ from $\Phi$.

%VVVVVVVVVVVVVVVVVVVVVVVVVVVVVVVVVVVVVVV
\subsection{Mediated interactions imply Debye shielding}
\label{MIIDS}
%VVVVVVVVVVVVVVVVVVVVVVVVVVVVVVVVVVVVVVV

In the above introduction of Debye shielding, using the Laplace transform of the particle positions does not provide an intuitive picture of this effect. We now show that such a picture can be obtained by turning back to the mechanical description of microscopic dynamics with the full Coulomb potential of Eq.~(\ref{phitildetotM}). In order to compute the dynamics, we use Picard iteration technique. From Eq.~(\ref{rsectot}), $\textbf{r}_l^{(n)}$, the $n$th iterate of $\textbf{r}_l$, is computed by
\begin{equation}
  \ddot{\textbf{r}}^{(n)}_l
  = \frac{e}{m_e} \nabla \varphi_l^{(n-1)}(\textbf{r}^{(n-1)}_l),
\label{rsecn}
\end{equation}
where $\varphi_l^{(n-1)}$ is computed by the inverse Fourier transform of Eq.~(\ref{phitildetotM}) with the $\textbf{r}_j$'s substituted with the $\textbf{r}_j^{(n-1)}$'s. The iteration starts with the ballistic approximation of the dynamics defined by Eq.~(\ref{rl0}), and the actual orbit of Eq.~(\ref{rsectot}) corresponds to $n \rightarrow \infty$. Let $\delta \textbf{r}_l^{(n)} = \textbf{r}_l^{(n)} - \textbf{r}_l^{(0)}$ be the mismatch of the position of particle $l$ with respect to the ballistic one at the $n$th iterate. It is convenient to write Eq.~(\ref{rsecn}) as
\begin{equation}
\delta \ddot{\textbf{r}}^{(n)}_l = \sum_{j \in S;j \neq l} \delta \ddot{\textbf{r}}^{(n)}_{lj},
\label{rsecnAccl}
\end{equation}
with
\begin{equation}
\delta \ddot{\textbf{r}}^{(n)}_{lj} = \textbf{a}_{\rm{C}}(\textbf{r}_l^{(n-1)}-\textbf{r}_j^{(n-1)}),
\label{rsecnAcclj}
\end{equation}
 and
\begin{equation}
\textbf{a}_{\rm{C}}(\textbf{r}) =  - \frac{e}{m_e} \nabla \delta \varphi_{\rm{C}} (\textbf{r}),
\label{AccC}
\end{equation}
with $\delta \varphi_{\rm{C}}$, the $L$-periodized, OCP bare Coulomb potential, given by
\begin{equation}
\delta \varphi_{\rm{C}}(\textbf{r}) =
 -\frac{e}{\varepsilon_0 L^3}
\sum_{{\textbf{m}}\neq {\textbf{0}}} k_{\textbf{m}}^{-2} \exp(\rmi \textbf{k}_{\textbf{m}} \cdot \textbf{r}),
\label{phiCbPer}
\end{equation}
such that $\varphi(\textbf{r}) = \sum_{j \in S} \delta \varphi_{\rm{C}}(\textbf{r} - \textbf{r}_j)$.
Then $\delta \ddot{\textbf{r}}^{(1)}_l = \sum_{j \in S;j \neq l} \textbf{a}_{\rm{C}}(\textbf{r}_l^{(0)}-\textbf{r}_j^{(0)})$ and for $n \geq 2$
\begin{equation}
\delta \ddot{\textbf{r}}^{(n)}_l = \delta \ddot{\textbf{r}}^{(1)}_l + \sum_{j \in S;j \neq l} \nabla \textbf{a}_{\rm{C}}(\textbf{r}_l^{(0)}-\textbf{r}_j^{(0)}) \cdot (\delta \textbf{r}_l^{(n-1)} - \delta \textbf{r}_j^{(n-1)}) + O(a^3),
\label{rsecnAccDev}
\end{equation}
where $a$ is the order of magnitude of the total Coulombian acceleration. Equation (\ref{rsecnAccDev}) may be written
\begin{equation}
\ddot{\textbf{r}}^{(n)}_l = \sum_{j \in S;j \neq l}
[(\delta \ddot{\textbf{r}}^{(1)}_{lj} + M_{lj}^{(n-1)})
+ 2 \nabla \textbf{a}_{\rm{C}}(\textbf{r}_l^{(0)}-\textbf{r}_j^{(0)}) \cdot \delta \textbf{r}_{lj}^{(n-1)}] + O(a^3),
\label{rsecnAccDev2}
\end{equation}
where $M_{lj}^{(n-1)} = \nabla \textbf{a}_{\rm{C}}(\textbf{r}_l^{(0)}-\textbf{r}_j^{(0)}) \cdot \sum_{i \in S; i \neq l,j} (\delta \textbf{r}_{li}^{(n-1)} - \delta \textbf{r}_{ji}^{(n-1)})$
is the modification of the bare Coulomb acceleration of particle $j$ on particle $l$ due to the following phenomenon: particle $j$ modifies the position of all other particles, which implies the action of the latter ones on particle $l$ is modified by particle $j$. Therefore  $M_{lj}^{(n-1)} $ is the acceleration of particle $l$ due to particle $j$ mediated by all other particles. The last term in the bracket in Eq.~(\ref{rsecnAccDev2}), where
$\delta \textbf{r}_{lj}^{(n-1)} = \int_0^t \int_0^{t'} \delta \ddot {\textbf{r}}_{lj}^{(n-1)}(t'') \rmd t'' \rmd t'$, accounts for the fact that both particles $j$ and $l$ are shifted with respect to their ballistic positions.

Since the shielded potential of section \ref{SCP} was found by first order perturbation theory, it is felt in the acceleration of particles computed to second order. This acceleration is provided by Eq.~(\ref{rsecnAccDev2}) for $n=2$. Therefore its term in brackets is the shielded acceleration of particle $l$ due to particle $j$. As a result, though the summation runs over all particles, its effective part is only due to particles $j$ typically inside the Debye sphere about particle $l$. Starting from the third iterate of Picard scheme, the effective part of the summation in Eq.~(\ref{rsecnAccDev2}) ranges inside this Debye sphere, since the $\delta \textbf{r}_{lj}^{(n-1)}$'s are then computed with a shielded acceleration. This justifies the use of the shielded potential to compute collisional transport in section \ref{DSCT}, and is the basis of the intuitive interpretation of Debye shielding in claim 9. This interpretation shows that, when starting form random particle positions, the typical time-scale for shielding to set in is the time for a thermal particle to cross a Debye sphere, i.e.\ $\omega_{\rmpp}^{-1}$, which sets the duration of the transients occurring in the inverse Laplace transform leading to Eq.~(\ref{phij}). This order of magnitude is correct for a plasma close to equilibrium.

%VVVVVVVVVVVVVVVVVVVVVVVVVVVVVVVVVVVVVVV
\subsection{Debye shielding and Landau damping}
\label{STSP}
%VVVVVVVVVVVVVVVVVVVVVVVVVVVVVVVVVVVVVVV

We now apply the smoothing using distribution function $f$ to $\tilde{\phi}^{(0)}(\textbf{m},\omega)$ too in Eq.~(\ref{phihatL}). As a result of Eqs~(\ref{phitildn}-\ref{phitildnj}), this yields
\begin{equation}
\tilde{\Phi}^{(0)}(\textbf{m})
  = - \frac{e}{\varepsilon_0 k_{\textbf{m}}^2}\iint  \exp[- \rmi \textbf{k}_{\textbf{m}} \cdot (\textbf{r} + \textbf{v} t)]f(\textbf{r},\textbf{v}) \rmd^3 \textbf{r} \rmd^3 \textbf{v}
\label{phi0tildetcg}
\end{equation}
whose Laplace transform is
\begin{equation}
\hat{\Phi}^{(0)}(\textbf{m},\omega)
  = - \frac{\rmi e}{\varepsilon_0 k_{\textbf{m}}^2} \int
\frac{\tilde{f}(\textbf{m},\textbf{v})}{\omega -\textbf{k}_{\textbf{m}} \cdot \textbf{v}} \ \rmd^3 \textbf{v},
\label{phi0hatcg}
\end{equation}
which shows this second smoothing makes Eq.~(\ref{phihatL}) to become the expression allowing Landau contour calculations, as stated in Claim 3.

We point out that, in this paper, the smoothed velocity distribution is introduced after particle dynamics has been taken into account, and not before, as occurs when kinetic equations are used. This avoids addressing the issues of the exact definition of the smoothed distribution for a given realization of the plasma, and of the uncertainty as to the way the smoothed dynamics departs from the actual $N$-body one.

%TTTTTTTTTTTTTTTTTTTTTTTTTTTTTTTTTTTTTT
\section{Wave-particle dynamics}
\label{WPDRT}
%TTTTTTTTTTTTTTTTTTTTTTTTTTTTTTTTTTTTTT

In section \ref{STSP}, the existence of Langmuir waves is asserted by connection with Landau theory. To describe Langmuir waves with discrete particles, we consider that the $\textbf{r}_{l0}$'s are random, and we allow for non zero $\delta \textbf{r}_j(0)$'s and $\delta \dot{\textbf{r}}_j(0)$'s for the $\delta \textbf{r}_j$'s in Eq.~(\ref{phitildnj}). Therefore, in the formulas of section \ref{DSLD}, the $\textbf{r}_{j0}$ and $\textbf{v}_j$'s are slightly shifted with respect to the initial $\textbf{r}_j(0)$'s and $\dot{\textbf{r}}_j(0)$'s due to Langmuir waves.

Up to this point we described Langmuir waves by a linear theory. We now generalize the analysis of section \ref{FEP} to afford the description of nonlinear effects in wave-particle dynamics. Indeed, resonant particles may experience trapping or chaotic dynamics, which imply $\textbf{k}_{\textbf{m}} \cdot \delta \textbf{r}_l$'s of the order of $2 \pi$ or larger for wave $\textbf{k}_{\textbf{m}}$'s. To describe such a dynamics, it is not appropriate to expand $\phi$ as was done in Eqs~(\ref{phitildn}-\ref{phitildnj}) for such particles. However, this expansion may still be justified for non resonant particles over times where trapping and chaos show up for resonant ones. In order to keep the capability to describe the latter effects, we now split the set of $N$ particles into bulk and tail, in the spirit of Refs. \cite{OWM,OLMSS,AEE,EZE,EEB}. The bulk is defined as the set of particles which are not resonant with Langmuir waves. We then perform the analysis of section \ref{FEP} for the $N_{\mathrm{bulk}}$ particles, while keeping the exact contribution of the $N_{\mathrm{tail}}$ particles to the electrostatic potential. To this end, we number the tail particles from 1 to $N_{\mathrm{tail}}$, the bulk ones from $N_{\mathrm{tail}}+1$ to $N = N_{\mathrm{bulk}} + N_{\mathrm{tail}}$, and we call these respective sets of integer $S_{\mathrm{tail}}$ and $S_{\mathrm{bulk}}$. For $l \in S_{\mathrm{bulk}}$, we now substitute Eq.~(\ref{phitildn}) with
\begin{equation}
\tilde{\phi}_l (\textbf{m}) = \frac{N_{\mathrm{bulk}}  - 1}{N_{\mathrm{bulk}}} U(\textbf{m})
+ \sum_{j \in S_{\mathrm{bulk}};j \neq l} \ \delta \tilde{\phi}_{j} (\textbf{m}) ,
\label{phitildnT}
\end{equation}
where
\begin{equation}
U(\textbf{m}) = -\frac{e N_{\mathrm{bulk}}}{\varepsilon_0 k_{\textbf{m}}^2 (N_{\mathrm{bulk}}-1)} \sum_{j \in S_{\mathrm{tail}}} \exp(- \rmi \textbf{k}_{\textbf{m}} \cdot \textbf{r}_j).
\label{U}
\end{equation}
In the r.h.s.\ of Eq.~(\ref{phitildnT}), the first term vanishes if $N_{\mathrm{tail}} = 0$. We now perform the calculation of section \ref{FEP} on substituting the previous summations with index running from 1 to $N$ by ones where the index runs over $S_{\mathrm{bulk}}$, while keeping the exclusion of $j=l$ where indicated. The previous division by $N-1$ preceding Eq.~(\ref{phihat}) is now a division by $ N_{\mathrm{bulk}} - 1$. This yields
\begin{eqnarray}
 k_{\textbf{m}}^2\hat{\varphi}(\textbf{m},\omega)
&-& \frac{e^2}{ L^3 m_e \varepsilon_0}
 \sum_{\textbf{n}} \textbf{k}_{\textbf{m}} \cdot \textbf{k}_{\textbf{n}}
\ \sum_{j \in S} \frac{\Psi_j(\hat{\varphi}\ ; \textbf{n},\omega + \omega_{\textbf{n},j} - \omega_{\textbf{m},j})}{(\omega - \omega_{\textbf{m},j})^2} \exp[\rmi (\textbf{k}_{\textbf{n}}-\textbf{k}_{\textbf{m}}) \cdot \textbf{r}_{j0}]
\nonumber\\
= k_{\textbf{m}}^2 \hat{\phi}^{(0)}(\textbf{m},\omega) &-& \frac{e}{\varepsilon_0} \sum_{j \in S} \exp(- \rmi \textbf{k}_{\textbf{m}} \cdot \textbf{r}_{j0}) \hat{R}_j(\omega - \omega_{\textbf{m},j})
+ k_{\textbf{m}}^2 \hat{U}(\textbf{m},\omega),
\label{phihatU}
\end{eqnarray}
where $\hat{U}(\textbf{m},\omega)$ is the Laplace transform of $U(\textbf{m})$. Then Eq.~(\ref{phihatL}) becomes
\begin{equation}
\epsilon(\textbf{m},\omega) \hat{\Phi}(\textbf{m},\omega) = \hat{\phi}^{(0)}(\textbf{m},\omega) + \hat{U}.
\label{phihatLU}
\end{equation}

Let $\tilde{\Phi}(\textbf{m},t)$ be the inverse Laplace transform of  $\hat{\Phi}(\textbf{m},\omega)$, $\hat{\Phi}_{\mathrm{bulk}}(\textbf{m},\omega)$ be the solution of Eq.~(\ref{phihatL}) computed for the bulk particles, and $\tilde{\Phi}_{\mathrm{bulk}}(\textbf{m},t)$ be its inverse Laplace transform. We now derive an amplitude equation for $\tilde{\Phi}(\textbf{m},t)$ in a way similar to Refs. \cite{OWM,OLMSS}. Let $\omega_{\textbf{m}}$ be such that $\epsilon(\textbf{m},\omega_{\textbf{m}}) = 0$; because of the definition of the bulk, this frequency is real. Then $\tilde{\Phi}_{\mathrm{bulk}}(\textbf{m},t) = A \exp(-\rmi \omega_{\textbf{m}} t)$, where $A$ is a constant, and
\begin{equation}
\hat{\phi}^{(0)}(\textbf{m},\omega) = \frac{\rmi A}{\omega - \omega_{\textbf{m}}},
\label{phi0A}
\end{equation}
according to Eq.~(\ref{phihatL}).

Let $g(\textbf{m},t) = \tilde{\Phi}( \textbf{m},t) / \tilde{\Phi}_{\mathrm{bulk}}(\textbf{m},t)$. Therefore $\hat{\Phi}(\textbf{m},\omega) = A \hat{g}(\omega - \omega_{\textbf{m}})$, which together with Eq.~(\ref{phihatLU}) and (\ref{phi0A}) yields
\begin{equation}
A \epsilon(\textbf{m},\omega_{\textbf{m}} + \omega') [\hat{g}(\textbf{m},\omega') - \frac{\rmi}{\omega'}] = \hat{U}(\textbf{m},\omega_{\textbf{m}} + \omega'),
\label{eqf}
\end{equation}
where $\omega' = \omega - \omega_{\textbf{m}}$. If $ N_{\mathrm{tail}} \ll N_{\mathrm{bulk}}$, $g(\textbf{m},t)$ is a slowly evolving amplitude, and the support of $\hat{g}(\textbf{m},\omega)$ is narrow about zero. This justifies Taylor-expanding $\epsilon(\textbf{m},\omega_{\textbf{m}} + \omega')$ about $\omega' = 0$ in Eq.~(\ref{eqf}), which yields $\frac{\partial \epsilon(\textbf{m},\omega_{\textbf{m}})}{\partial \omega} \omega'$ to lowest order. Setting this in Eq.~(\ref{eqf}) and performing the inverse Laplace transform finally yields an amplitude equation for $\tilde{\Phi}(\textbf{m},t)$
\begin{equation}
\frac{\partial \tilde{\Phi}(\textbf{m},t)}{\partial t}  + \rmi \omega_{\textbf{m}} \tilde{\Phi}(\textbf{m},t) =
 \frac{\rmi e N_{\mathrm{bulk}}}{\varepsilon_0 k_{\textbf{m}}^2 (N_{\mathrm{bulk}}  - 1) \frac{\partial \epsilon(\textbf{m},\omega_{\textbf{m}})}{\partial \omega}} \sum_{j \in S_{\mathrm{tail}}} \exp(- \rmi \textbf{k}_{\textbf{m}} \cdot \textbf{r}_j).
\label{eqampl}
\end{equation}
The self-consistent dynamics of the potential and of the tail particles is ruled by this equation and by the equation of motion of these particles
\begin{equation}
\ddot{\textbf{r}}_j = \frac{\rmi e}{L^3 m_e} \sum_{\textbf{n}} \textbf{k}_{\textbf{n}} \ \tilde{\Phi}_j(\textbf{n}) \exp(\rmi \textbf{k}_{\textbf{n}} \cdot \textbf{r}_j).
\label{delrsecwv}
\end{equation}

These two sets of equations generalize to 3 dimensions the self-consistent dynamics defined in Refs. \cite{AEE,EEB}. For the sake of brevity, we do not develop here the full generalization of the analysis in these papers; it is lengthy, but straightforward. However, since this analysis unifies spontaneous emission and Landau growth and damping, we give the result ruling the evolution of the amplitude of a Langmuir wave provided by perturbation calculation where the right hand sides of Eqs~(\ref{eqampl}-\ref{delrsecwv}) are considered as small of order one. This is natural for Eq.~(\ref{eqampl}) since $N_{\mathrm{tail}} \ll N_{\mathrm{bulk}}$, and for Eq.~(\ref{delrsecwv}) if the Langmuir waves have a low amplitude. Let $J(\textbf{m},t) = \langle \tilde{\Phi}(\textbf{m},t)\tilde{\Phi}(- \textbf{m},t) \rangle$, where the average is over the random initial positions of the tail particles (thus their distribution is spatially uniform). Then a second order calculation in $\Phi$ yields \begin{equation}
    \frac {\rmd J(\textbf{m},t)} {\rmd t}
    = 2 \gamma_{\textbf{m}{\rm L}} J(\textbf{m},t) + S_{\textbf{m} \, \mathrm {\rm{spont}} },
   \label{evampfinal}
\end{equation}
where $\gamma_{\textbf{m}{\rm L}}$ is the Landau growth or damping rate given by
\begin{equation}
  \gamma_{\textbf{m} {\rm L}} = \alpha_{\textbf{m}}
  {\frac {\rmd f_{\rm{red}}} {\rmd v}}(\frac{\omega_{\textbf{m}}}{ k_{\textbf{m}}} ; \textbf{m})
  \label{SZ122}
\end{equation}
with
\begin{equation}
  \alpha_{\textbf{m}}
  = \frac{\pi e^2 }{m_e \varepsilon_0 k_{\textbf{m}}^2  \frac{\partial \epsilon(\textbf{m},\omega_{\textbf{m}})}{\partial \omega}},
 \label{alj}
\end{equation}
and $f_{\rm{red}}(v; \textbf{m})$ is the reduced smoothed distribution function $f_{\rm{red}}(v; \textbf{m}) = \iint f(v \hat{\textbf{k}}_{\textbf{m}} + \textbf{v}_{\bot}) \ \rmd^2 \textbf{v}_{\bot}$ where $\hat{\textbf{k}}_{\textbf{m}}$ is the unit vector along $\textbf{k}_{\textbf{m}}$ and $\textbf{v}_{\bot}$ is the component of the velocity perpendicular to $\textbf{k}_{\textbf{m}}$; $S_{\textbf{m} \, \mathrm {\rm{spont}} }$ is given by
\begin{equation}
  S_{\textbf{m} \, \mathrm {\rm{spont}} } = \frac{2 \alpha_{\textbf{m}}^2}{\pi e^2  k_{\textbf{m}} n} f_{\rm{red}}(\frac{\omega_{\textbf{m}}}{ k_{\textbf{m}}}),
  \label{Spont}
\end{equation}
where $ n=N/L^3$ is the plasma density. $S_{\textbf{m} \, \mathrm {\rm{spont}} }$ corresponds to the spontaneous emission of waves by particles and induces an exponential relaxation of the waves to the thermal level in the case of Landau damping. The second order calculation for the particles yields the diffusion and friction coefficients of the Fokker-Planck equation ruling the tail dynamics. This equation corresponds to the classical quasilinear result, plus a dynamical friction term mirroring the spontaneous emission of waves by particles, as found in the one-dimensional case in Refs. \cite{EZE,EEB}.

Finally, we point out that Eqs~(\ref{eqampl}-\ref{delrsecwv}) enable to tackle the linear theory of Langmuir waves by considering a single mechanical system \cite{EZE,EEB}. This is naturally obtained by considering the unperturbed plasma as made up of a series of monokinetic parallel beams, and each beam as an array of equidistributed particles. This approach finds eigenmodes corresponding to the Landau instability \cite{La}, if the distribution function has a positive slope. If the slope is negative, it recovers the van Kampen modes \cite{vK}, as a series of beam modes providing Landau damping as a result of the phase mixing of the latter. This phase mixing turns out to be also important to recover a correct growth of the wave amplitude in the unstable case (section 3.8.3 of reference \cite{EEB}). We point out that in Refs. \cite{EZE,EEB} the equivalent of Eqs~(\ref{eqampl},\ref{delrsecwv}) was obtained without using any smoothing, but by a mechanical reduction of degrees of freedom, which made even stronger the idea of considering a single mechanical system.

%ZZZZZZZZZZZZZZZZZZZZZZZZZZZZZZZZZ
\section{Debye shielding and collisional transport}
\label{DSCT}
%ZZZZZZZZZZZZZZZZZZZZZZZZZZZZZZZZZ

We now focus on the case where the particles have random initial positions, i.e.\ where the plasma has a uniform density, and for simplicity we consider the plasma to be in thermal equilibrium. Then the dynamics of particles has no collective aspect, but is ruled by the cumulative effect of two-body deflections. More specifically, we choose random $\textbf{r}_{l0}$'s, and vanishing $\delta \textbf{r}_l(0)$'s and $\delta \dot{\textbf{r}}_l(0)$'s, and we assume that at $t=0$ all particles are in the same cube with volume $L^3$, and we consider the limit $L/\lambda_{\mathrm{D}} \rightarrow \infty$. In contrast, each particle has a well defined velocity, in such a way that the overall initial smoothed velocity distribution is close to some given function. We focus on particle $l$ which is assumed to be close to the center of the cube. In this section, we approximate the true dynamics by that due to the shielded Coulombian interactions, i.e.\ we write
\begin{equation}
\delta \ddot{\textbf{r}}_l = \sum_{j \in S;j \neq l} \textbf{a}(\textbf{r}_l-\textbf{r}_j,\textbf{v}_j),
\label{delrsecscreen}
\end{equation}
with
\begin{equation}
\textbf{a}(\textbf{r},\textbf{v}) = \frac{e}{m_e} \nabla \delta \Phi (\textbf{r},\textbf{v}),
\label{acc}
\end{equation}
where $\delta \Phi (\textbf{r},\textbf{v})$ is given by Eq.~(\ref{phi}).

We compute particle $l$ deflection in a sequence of steps. First, we use first order perturbation theory in $\delta \Phi$, which shows the total deflection to be the sum of the individual deflections due to all other particles. For an impact parameter much smaller than $\lambda_{\rm{D}}$, the deflection due to a particle turns out to be the perturbative value of the Rutherford deflection due to this particle if it were alone. Second, for a close encounter with particle $n$, we show that the deflection of particle $l$ is exactly the one it would undergo if the other $N-2$ particles were absent. Third, the deflection for an impact parameter of order $\lambda_{\rm{D}}$ is given by the Rutherford expression multiplied by some function of the impact parameter reflecting shielding. These three steps yield an analytical expression for deflection whatever the impact parameter.

We first compute $\delta \textbf{r}_l$ by first order perturbation theory in $\delta \Phi$, taking the ballistic motion defined by Eq.~(\ref{rl0}) as zeroth order approximation. This yields
\begin{equation}
\delta \dot{\textbf{r}}_{l1}(t) = \sum_{j \in S;j \neq l} \delta \dot{\textbf{r}}_{lj1}(0,t) ,
\label{delrl1}
\end{equation}
where
\begin{equation}
\delta \dot{\textbf{r}}_{lj1}(t_1,t_2) = \int_{t_1}^{t_2}  \textbf{a}[\textbf{r}_l^{(0)}(t')-\textbf{r}_j^{(0)}(t'),\textbf{v}_j] \rmd t'.
\label{delrlj1}
\end{equation}
 It is convenient to write
 \begin{equation}
 \textbf{r}_l^{(0)}(t')-\textbf{r}_j^{(0)}(t') = \textbf{b}_{lj}+ \Delta \textbf{v}_{lj} (t' - t_{lj}),
  \label{blj}
\end{equation}
where $t_{lj}$ is the time of closest approach of the two ballistic orbits, and $\textbf{b}_{lj}$ is the vector joining particle $j$ to particle $l$ at this time. Then $b_{lj} = \| \textbf{b}_{lj}\|$ is the impact parameter of these two orbits when singled out. The initial random positions of the particles translate into random values of $\textbf{b}_{lj}$ and of $t_{lj}$. The typical duration of the deflection of particle $l$ given by Eq.~(\ref{delrlj1}) is $\Delta t_{lj} \equiv b_{lj} / \Delta v_{lj}$ where $\Delta v_{lj} = \| \Delta \textbf{v}_{lj} \|$, but a certain number, say $\alpha$, of $\Delta t_{lj}$'s are necessary for the deflection to be mostly completed. For a given $b_{lj} $ and for $t \gg \Delta t_{lj}$, the deflection of particle $l$ given by Eq.~(\ref{delrlj1}) is maximum if $t_{lj}$ is in the interval $[\alpha \Delta t_{lj}, t - \alpha \Delta t_{lj}]$. We notice that $\Delta t_{lj}$ is about the inverse of the plasma frequency for $b_{lj} \sim \lambda_{\rm{D}}$ and $\Delta v_{lj}$ on the order of the thermal velocity.

For the sake of brevity, we compute here just the trace of the diffusion tensor for the particle velocities. To this end, we perform an average over all the $\textbf{r}_{l0}$'s to get
\begin{equation}
\langle \delta \dot{\textbf{r}}_{l1}^2(t)\rangle = \sum_{j \in S;j \neq l}
\langle \delta \dot{\textbf{r}}_{lj1}^2(t)\rangle,
\label{delrlj1^2}
\end{equation}
taking into account Eq.~(\ref{phi}), and the fact that the initial positions are independently random, as well as the $\textbf{r}_i - \textbf{r}_j$'s for $i \neq j$. Therefore, though being due to the simultaneous scattering of particle $l$ with the many particles inside its Debye sphere, $\langle \delta \dot{\textbf{r}}_{l1}^2(t)\rangle$ turns out to be the sum of individual two-body deflections for $b_{lj} $'s such that first order perturbation theory is correct. Hence the contribution to $\langle \delta \dot{\textbf{r}}_{l1}^2(t)\rangle$ of particles with given $b_{lj}$ and $\Delta v_{lj}$ can be computed as if it would result from successive two-body collisions, as was done in Ref.\ \cite{Ros} and in many textbooks.

For an impact parameter much smaller than $\lambda_{\rm{D}}$, the main contribution of $\textbf{a}[\textbf{r}_l^{(0)}(t')-\textbf{r}_j^{(0)}(t'),\textbf{v}_j]$ to the deflection of particle $l$ comes from times $t'$ where $\| \textbf{r}_l^{(0)}(t')-\textbf{r}_j^{(0)}(t') \| \ll \lambda_{\rm{D}}$. Therefore $\textbf{a}(\textbf{r},\textbf{v})$ takes on its bare Coulombian value, and $\langle \delta \dot{\textbf{r}}_{l1}^2(t)\rangle$ is a first order approximation of the effect on particle $l$ of a Rutherford collision with particle $j$. Comparing this approximate value with the exact one shows the perturbative calculation to be correct for $b_{lj} \gg \lambda_{\rm{ma}} = \frac{e^2}{ \pi m_e \varepsilon_0 \Delta v_{lj}^2}$, the distance of minimum approach of two electrons in a Rutherford collision, as given by energy conservation.

Second, we consider the case of the close approach of particle $n$ to particle $l$, i.e.\ $b_{ln} \sim \lambda_{\rm{ma}}$. We write the acceleration of particle $l$ as
\begin{equation}
\ddot{\textbf{r}}_l = \textbf{a}(\textbf{r}_l-\textbf{r}_n) + \sum_{j \in S;j \neq l,n} \textbf{a}(\textbf{r}_l-\textbf{r}_j).
\label{delrsecEx}
\end{equation}
For particle $n$, we write the same equation by exchanging indices $l$ and $n$. Since the two particles are at distances much smaller than the inter-particle distance $d = n^{-1/3} = L/N^{1/3}$, the accelerations they get from all other particles are almost the same. Therefore, when subtracting the two rigorous equations of motion, the two summations over $j$ almost cancel, yielding

\begin{equation}
\frac{\rmd^2 (\textbf{r}_l - \textbf{r}_n)}{\rmd t^2} = 2 \textbf{a}(\textbf{r}_l-\textbf{r}_n),
\label{delrsecEx2}
\end{equation}
which is the equation describing the Rutherford collision of these two particles in their center of mass frame, in the absence of all other particles (at such distances the shielded potential is the bare Coulomb one). Since $b_{ln} \ll d$, $\Delta t_{ln}$ is much smaller than the $\Delta t_{lj}$'s of the other particles. Therefore the latter produce a negligible deflection of the center of mass during the Rutherford two-body collision, and the deflection of particle $l$ during this collision is exactly that of a Rutherford two-body collision. The contribution of such collisions to $\langle \delta \dot{\textbf{r}}_{l}^2(t)\rangle $ was calculated in Ref.\ \cite{Ros}.

Now, since the deflection of particle $l$ due to particle $j$ as computed by the above perturbation theory is an approximation of the Rutherford deflection for the same impact parameter, we may approximate the perturbative deflection one by the full Rutherford one, which provides an obvious matching of the theories for $b_{lj} \sim \lambda_{\rm{ma}}$ and for $\lambda_{\rm{D}} \gg b_{lj} \gg \lambda_{\rm{ma}}$: we may use the estimate of \cite{Ros} in the whole domain $b_{lj} \ll \lambda_{\rm{D}}$.

Third, we deal with impact parameters of the order of $\lambda_{\rm{D}}$. Then the deflection due to particle $j$ must be computed with Eq.~(\ref{delrlj1}). For the sake of simplicity, we make the calculation for the case where $\textbf{v}_j$ is small, which makes $\delta \Phi (\textbf{r},\textbf{v}) \simeq \delta \Phi (\textbf{r},\textbf{0})$ which is the Yukawa potential $\delta \Phi_{\rm{Y}} (\textbf{r}) = - \frac{e}{4 \pi \varepsilon_0 \| \textbf{r} \|} \exp (- \frac{\| \textbf{r} \|}{\lambda_{\rm{D}}})$ (Eq.~(18) of Ref.\ \cite{Gasio}). The first order correction in $\textbf{k}_{\textbf{m}}  \cdot \textbf{v}_j$ to this approximation is a dipolar potential with an electric dipole moment proportional to $\textbf{v}_j$. Since a Maxwellian distribution is symmetrical in $\textbf{v}$, these individual dipolar contributions cancel globally. As a result, the first relevant correction to the Yukawa potential is of second order in $\textbf{k}_{\textbf{m}}  \cdot \textbf{v}_j$. This should make the Yukawa approximation relevant for a large part of the bulk of the Maxwellian distribution.

In the small deflection limit, a calculation using the fact that the force derives from a central potential shows the full deflection of particle $l$ due to particle $j$ is provided by
\begin{equation}
\delta \dot{\textbf{r}}_{lj1}(- \infty,+ \infty) =
\frac{e^2}{ 4 \pi m_e \varepsilon_0} \textbf{b}_{lj}
\int_{- \infty}^{+ \infty}  [\frac{1}{r^3(t)} + \frac{1}{\lambda_{\rm{D}} r^2(t)}] \exp [- \frac{r(t)}{\lambda_{\rm{D}}}]\rmd t,
\label{delrljT}
\end{equation}
where $r(t) = (b_{lj}^2 +\Delta v_{lj}^2 t^2)^{1/2}$ and $\textbf{b}_{lj}$ was defined with Eq.~(\ref{blj}). Defining $\theta = \arcsin [ \Delta v_{lj} t / r(t)]$, this equation becomes
\begin{equation}
\delta \dot{\textbf{r}}_{lj1}(- \infty,+ \infty) =
- \frac{2 e^2}{ 4 \pi m_e \varepsilon_0 \Delta v_{lj}}
 \, \frac{h(b_{lj})}{b_{lj}^2 } \, \textbf{b}_{lj},
\label{delrljTfin}
\end{equation}
where
\begin{equation}
h(b) =
\int_{0}^{\pi/2} [ \cos (\theta) +  \frac{b}{\lambda_{\rm{D}}}] \exp [- \frac{b}{\lambda_{\rm{D}} \cos (\theta)}] \ \rmd \theta
< [1 + \frac{\pi b}{2 \lambda_{\rm{D}}}] \exp [- \frac{b}{\lambda_{\rm{D}}}].
\label{delrljT2}
\end{equation}
During time $t \gg \Delta t_{lj}$, a volume $2 \pi \Delta v_{lj} t b_{lj} \delta b_{lj}$ of particles with velocity $\textbf{v}_j$ and impact parameters between $b_{lj}$ and $b_{lj} + \delta b_{lj}$ produce the deflection of particle $l$ given by Eq.~(\ref{delrljTfin}), and a contribution scaling like $\frac{h^2(b_{lj})}{b_{lj}} \delta b_{lj}$ to $\langle \delta \dot{\textbf{r}}_{l1}^2(t)\rangle$. Let $b_{\rm{min}}$ be such that $\lambda_{\rm{D}} \gg b_{\rm{min}} \gg \lambda_{\rm{ma}}$. The contribution of all impact parameters between $b_{\rm{min}}$ and some $b_{\rm{max}}$ is thus scaling like the integral $\int_{b_{\rm{min}}}^{b_{\rm{max}}} h^2(b)/b  \ \rmd b$. Since for $b$ small $h(0) \simeq 1$, if $b_{\rm{max}} \ll \lambda_{\rm{D}}$, this is the non-shielded contribution of orbits relevant to the above perturbative calculation. Since, by approximating it by the Rutherford-like result of Ref.\ \cite{Ros}, this contribution matches that for impact parameters on the order of $\lambda_{\rm{ma}}$, the contribution of all impact parameters between $\lambda_{\rm{ma}}$ and some $b_{\rm{max}}$ small with respect to $\lambda_{\rm{D}}$ is thus scaling like the integral $\int_{\lambda_{\rm{ma}}}^{b_{\rm{max}}} 1/b  \ \rmd b$ as was computed in Ref.\ \cite{Ros}. The matching of this result for $b \sim \lambda_{\rm{D}}$ is simply accomplished by setting a factor $h^2(b)$ in the integrand which makes the integral converge for $b \rightarrow \infty$. Taking this limit, one finds that the Coulomb logarithm $\ln (\lambda_{\rm{D}} / \lambda_{\rm{ma}})$ of the second Eq.~(14) of Ref.\ \cite{Ros} becomes $\ln (\lambda_{\rm{D}} / \lambda_{\rm{ma}}) + C$ where $C$ is of order unity. If the full dependence of the shielding on $\textbf{v}_j$ were taken into account, the modification of the Coulomb logarithm would be velocity dependent.

For the sake of brevity, we do not develop here the calculation of the dynamical friction which will be presented elsewhere. This calculation requires using second order perturbation theory, but follows the same lines as those for the diffusion coefficient.

For an inhomogeneous plasma, the acceleration of particle $l$ may be split into a homogeneous and a wave part. When using the linearized versions of Eq.~(\ref{phihat}) and subsequent ones in section \ref{DSLD}, we can split all $\phi$'s and $\Phi$'s into a homogeneous part and an independent inhomogeneous one. Therefore the diffusion coefficient and the dynamical friction estimated by perturbative calculation of the dynamics up to second order are the sum of the collisional contribution and of a contribution due to waves, the latter as calculated for instance in Refs. \cite{EZE,EEB}. For the sake of brevity, we defer this point to later publication.

%TTTTTTTTTTTTTTTTTTTTTTTTTTTTTTTTTTTTTT
\section{Conclusion}
\label{Concl}
%TTTTTTTTTTTTTTTTTTTTTTTTTTTTTTTTTTTTTT

In this paper, Debye shielding, collisional transport, Landau damping of Langmuir waves, and spontaneous emission of these waves were introduced by a mechanical approach. This approach brings unification and simplification in basic microscopic plasma physics, and may be useful for pedagogical purposes. One might think about trying to apply the above approach to plasmas with more species, or with a magnetic field, or where particles experience trapping and chaotic dynamics. The first generalization sounds rather trivial, and the third one is under way, at least in one dimension (see a pedestrian introduction in \cite{Houches} and more specific results in \cite{BEEB,BEEBEPS}).

As in many textbooks, linearization was applied in sections \ref{FEP}, \ref{SCP}, and \ref{STSP} without questioning deeply its range of validity. However, the smallness of the perturbation is not a sufficient criterion. Indeed, as reviewed in Ref.\ \cite{HS}, perturbation theory that relies on linearization has to be questioned, as it yields a solution of the linearized set of equations only. Whether it also generates a solution of the full set has to be shown explicitly, and this may be a hard task. For instance the full proof of existence of Landau damping \cite{MV} in a Vlasovian frame was a mathematical tour de force, the equivalent of a Kolmogorov-Arnold-Moser theorem for continuous systems, and led one of its authors to be awarded the 2010 Fields medal.

As to collisional transport, as usual, it was assumed that some kind of decorrelation justifies the calculation of the transport coefficients over the finite times considered in section \ref{DSCT}. However, in reality particle dynamics is chaotic, and one is facing the calculation of transport coefficients for a chaotic motion. One should be cautious as, for the motion of a charged particle in a spectrum of longitudinal waves, a perturbation calculation yields the quasilinear estimate for the diffusion coefficient,
while a super-quasilinear regime, a synergetic effect in chaos, is found to exist in this chaotic dynamics for intermediate resonance overlap \cite{CEV}. There, diffusion becomes quasilinear for strong resonance overlap, but not because the perturbation calculation becomes valid again (\cite{EE2}, section 6.8.2 of \cite{EEB}).

Furthermore, we used only a very specific part of the fundamental equation (\ref{phihat}): the one involving linearization and smoothing. It would be interesting to study the effect of the coupling of Fourier components with both coherent and incoherent effects, in particular, to perform the analysis of section \ref{WPDRT} by substituting $k_{\textbf{m}}^2 \hat{U}(\textbf{m},\omega)$ with $- \frac{e}{\varepsilon_0} \sum_{j \in S} \exp(- \rmi \textbf{k}_{\textbf{m}} \cdot \textbf{r}_{j0}) \hat{R}_j(\omega - \omega_{\textbf{m},j})$. The question arises: is it possible to recover the hole solutions propagating near thermal velocity or slower, which are smooth and nonlinear structures satisfying the full nonlinear Vlasov-Poisson system (Ref.\ \cite{HS} and references therein)?
\vspace{8 mm}
%TTTTTTTTTTTTTTTTTTTTTTTTTTTTTTTTTTTTTT

Ph. Choquard, L. Cou\"edel, M.-C. Firpo, W. Horton, J.T. Mendon\c{c}a, F. Pegoraro, Y. Peysson, H. Schamel, and J.-Zh. Zhu are thanked for very useful comments and new references.

%TTTTTTTTTTTTTTTTTTTTTTTTTTTTTTTTTTTTTT

\end{document}